\documentclass[a4paper,12pt,english]{article}
\usepackage{amsfonts}
\usepackage{graphicx}
\usepackage{amsmath}
\usepackage{color}

\newcommand{\be}{\begin{equation}}
\newcommand{\ee}{\end{equation}}
\input{tcilatex}

\begin{document}


\begin{titlepage}
\begin{center}

\noindent{{\LARGE{Violating the string winding number maximally in Anti-de Sitter space}}}

\smallskip
\smallskip

\smallskip
\smallskip
\smallskip
\smallskip
\noindent{\large{Gaston Giribet}}

\smallskip
\smallskip

\end{center}
\smallskip
\smallskip
\centerline{Departamento de F\'{\i}sica, Universidad de Buenos Aires FCEN-UBA}
\centerline{{\it Ciudad Universitaria, Pabell\'on 1, 1428, Buenos Aires, Argentina.}}
\smallskip
\smallskip
\smallskip
\smallskip
\centerline{Instituto de F\'{\i}sica de Buenos Aires CONICET}
\centerline{{\it Ciudad Universitaria, Pabell\'on 1, 1428, Buenos Aires, Argentina.}}

\bigskip

\bigskip

\bigskip

\bigskip

\begin{abstract}
We study $n$-string scattering amplitudes in three-dimensional Anti-de Sitter space (AdS$_3$). 
We focus our attention on the processes in which the winding number conservation is violated maximally;
that is, those processes in which it is violated in $n-2$ units.
A worldsheet conformal field theory calculation leads us to confirm a previous conjecture about the functional
form of these observables.

\end{abstract}

\end{titlepage}


\section{Introduction}

String theory in AdS$_{3}$ has served as a model to explore AdS/CFT
correspondence beyond the supergravity approximation. What makes this
possible is that in the three-dimensional case one has access to the
worldsheet conformal field theory formulation in terms of the $SL(2,\mathbb{R%
})$ Wess-Zumino-Witten model (WZW) \cite{MO1, MO3}, and thus several
observables of the theory, like three-point functions, can be solved
exactly. This, together with the existence of a non-renormalization argument 
\cite{deBoer}, has led the authors of \cite{Gaberdiel, Pakman1, Pakman2} to
perform explicit checks of the matching between bulk and boundary
correlators at the string theory level (see also \cite{Rastelo, Carmen,
Kirsch1, Kirsch2}). The agreement found was exact, and this represents one
of the few examples in which one sees the bulk-boundary correspondence to
hold beyond the field theory approximation.

Besides its correlators being solvable, string theory in AdS$_{3}$ presents
several interesting aspects. In particular, its spectrum is very rich and
exhibits intriguing properties. As observed in \cite{MO1, MO3}, in order to
completely parameterize the spectrum of string theory in AdS$_{3}$, it is
necessary to consider spectrally flowed sectors of the space of
representation of the $sl(2)_{k}$ affine algebra. These sectors are labeled
by an integer number $\omega $, whose physical interpretation is that of
specifying the winding number of the string states. This winding number is a
dynamical degree of freedom associated to the presence of the $B$-field in
the background. Not being a topological quantity, this winding number can in
principle be violated if a string scattering process takes place. However,
such a violation is not totally arbitrary, and it happens to be bounded from
above following a curious pattern: In a scattering process that involves $n$
strings, the total winding number $\Delta \omega
=\tsum\nolimits_{i=1}^{n}\omega _{i}$ is restricted to obey the bound $%
|\Delta \omega |\leq n-2$; see \cite{MO3}. In this paper, we will focus on
the case where this bound is saturated; namely we will analyze $n$-string
interaction processes satisfying $|\Delta \omega |=n-2$.

$n$-string amplitudes in AdS$_{3}$ space admit to be written in terms of
Liouville theory correlation functions. This was explicitly shown in \cite{R}
for the case in which the winding number conservation is violated up to $n-3$
units (i.e. $|\Delta \omega |<n-2$.) A natural expression for the maximal
case $|\Delta \omega |=n-2$ was also proposed in \cite{R} following an
educated guess; however, in such special case the proof in \cite{R}, based
on the analysis of modular differential equations \cite{RibaultTeschner,
Stoyanovsky}, does not hold because the corresponding Liouville $n$-point
correlators do not generically involve degenerate primaries (see also \cite%
{HikidaSchomerus}). Therefore, when the winding number conservation is
violated maximally, the formula in \cite{R}\ that expresses $n$-string
amplitudes in AdS$_{3}$ in terms of $n$-point Liouville correlators remains
a conjecture. The aim of this paper is to show that a free field computation
in the worldsheet conformal field theory actually confirms that formula.
Proving so, amounts to review the free field formalism introduced in
references \cite{GN2, GN3}, which is particularly useful to compute
worldsheet correlation functions that violate the winding number
conservation. The formalism is based on the Dotsenko conjugate
representations introduced for the $SU(2)$ case in references \cite%
{Dotsenko1, Dotsenko2} and extended to the $SL(2,\mathbb{R})$ case in
references \cite{GN2, GN3, GiribetLopez}.

\section{Worldsheet conformal field theory in AdS$_{3}$}

Let us begin by briefly reviewing string theory in AdS$_{3}$. The theory is
described by the level-$k$ $SL(2,\mathbb{R})$ WZW model, where $k$ is given
by $k=l^{2}/\alpha ^{\prime }$, being $-l^{-2}$ the curvature of AdS$_{3}$.
The string spectrum is given by a subset of the direct sum of unitary $SL(2,%
\mathbb{R})$ representations \cite{MO1}; while discrete representations
correspond to short string states, the continuous series correspond to long
string states, for which we do have an S-matrix interpretation. The string
scattering amplitudes in AdS$_{3}$ are then given by integrating the $SL(2,%
\mathbb{R})$ WZW correlation functions over the worldsheet \cite{MO3}.

Correlation functions in the $SL(2,\mathbb{R})$ WZW model are defined by
analytic continuation of correlation functions in the model formulated on $%
\mathbb{H}_{+}^{3}$, which corresponds to Euclidean AdS$_{3}$. It is
convenient to start by discussing the model on $\mathbb{H}_{+}^{3}$ first.
More specifically, we should actually start by considering the model on $%
\mathbb{H}_{+}^{3}/U(1)\times \mathbb{R}$, which, as suggested in \cite{MO1}
and shown in \cite{GN2, GN3, GiribetLopez}, is the adequate construction to
describe winding string states. To describe the model on $\mathbb{H}%
_{+}^{3}/U(1)\times \mathbb{R}$, one may use Wakimoto free field
representation \cite{Wakimoto} with the addition of extra fields: First, one
adds a spacelike $U(1)$ free boson $X(z)$ to realize the coset $\mathbb{H}%
_{+}^{3}/U(1)$, as in \cite{BK,Becker}, and then adds an extra timelike free
boson $T(z)$ to represent the $\mathbb{R}$ time direction. The piece $%
\mathbb{H}_{+}^{3}$ is realized by the standard Wakimoto representation,
which consists of a $\beta $-$\gamma $ ghost system and a boson $\phi $ with
background charge.

Being described by the WZW model, the theory exhibits $SL(2,\mathbb{R}%
)_{k}\times SL(2,\mathbb{R})_{k}$ affine Kac-Moody symmetry, whose
holomorphic part can be expressed in terms of the following operator product
expansion (OPE)%
\begin{eqnarray}
J^{3}(z)J^{\pm }(w) &\simeq &\pm \frac{J^{\pm }(w)}{(z-w)}+\mathcal{O}(1) \\
J^{+}(z)J^{-}(w) &\simeq &\frac{k}{(z-w)^{2}}+\frac{2\ J^{3}(w)}{(z-w)}+%
\mathcal{O}(1) \\
J^{3}(z)J^{3}(w) &\simeq &\frac{-k/2}{(z-w)^{2}}+\mathcal{O}(1)
\end{eqnarray}%
where the $\mathcal{O}(1)$ stand for regular terms. Analogous OPE holds for
the anti-holomorphic piece. The double poles in the OPE above encode the
contribution of the central element of the $sl(2)_{k}$ affine algebra.

Using the free field representation mentioned above, the $sl(2)_{k}$
currents may be realized as follows%
\begin{eqnarray}
J^{+}(z) &=&\beta (z)\ e^{i\sqrt{\frac{2}{k}}(X(z)+T(z))}, \\
J^{3}(z) &=&-\beta (z)\gamma (z)-\sqrt{\frac{k-2}{2}}\partial \phi (z)-i%
\sqrt{\frac{k}{2}}\partial X(z)-i\sqrt{\frac{k}{2}}\partial T(z), \\
J^{-}(z) &=&(\beta (z)\gamma ^{2}(z)+\sqrt{2k-4}\gamma (z)\partial \phi
(z)+k\partial \gamma (z))\ e^{-i\sqrt{\frac{2}{k}}(X(z)+T(z))}.
\end{eqnarray}%
with the free field propagators%
\begin{equation}
\left\langle \phi (z)\phi (w)\right\rangle =\left\langle
X(z)X(w)\right\rangle =-\left\langle T(z)T(w)\right\rangle =-\log (z-w)\text{%
,\quad }\left\langle \beta (z)\gamma (w)\right\rangle =\frac{1}{(z-w)}\text{;%
}  \label{propagators}
\end{equation}%
and analogously for the anti-holomorphic contributions.

The states of the theory are labeled by indices $j$ and $m$, as it is usual
when classifying representations of $SL(2,\mathbb{R})$. It is also necessary
to introduce an integer index $\omega $ to specify which spectral flow
sector of the $sl(2)_{k}$ algebra the states are Kac-Moody primary with
respect to. Then, we denote the states by kets $\left\vert j,m,\overline{m}%
,\omega \right\rangle $.

The vertex operators that create these states are%
\begin{equation}
\Phi _{j,m,\overline{m}}^{\omega }(z)=c_{0}\ \gamma _{(z)}^{j-m}e^{\sqrt{%
\frac{2}{k-2}}j\phi (z)}e^{i\sqrt{\frac{2}{k}}mX(z)}e^{i\sqrt{\frac{2}{k}}(m+%
\frac{k}{2}\omega )T(z)}\times h.c.  \label{vertex}
\end{equation}%
where $h.c.$ stands for Hermitian conjugate, which is actually a misnomer as
it involves the contributions that depend on $\overline{m}$. The factor $%
c_{0}$ is a normalization constant, independent of $j$ and $m$.

Operators (\ref{vertex}) create the \textit{in}-states from the $SL(2,%
\mathbb{R})$ invariant vacuum $\left\vert 0\right\rangle $; namely%
\begin{equation}
\lim_{z\rightarrow 0}\Phi _{j,m,\overline{m}}^{\omega }(z)\left\vert
0\right\rangle =\left\vert j,m,\overline{m},\omega \right\rangle  \label{in}
\end{equation}%
as well as the \textit{out}-states%
\begin{equation}
\left\langle j,m,\overline{m},\omega \right\vert =\lim_{z\rightarrow \infty }%
\text{ }z^{2h_{j,m}^{\omega }}\overline{z}^{2h_{j,\overline{m}}^{\omega }}%
\text{ }\left\langle 0\right\vert \Phi _{-1-j,m,\overline{m}}^{\omega }(z),
\label{out}
\end{equation}%
where $h_{j,m}^{\omega }$ is the conformal dimension of the operators,%
\begin{equation}
h_{j,m}^{\omega }=-\frac{j(j+1)}{k-2}-m\omega -\frac{k}{4}\omega ^{2}.
\label{h}
\end{equation}

It is worth noticing that the formula for the conformal dimension remains
unchanged under the Weyl reflection $j\rightarrow -1-j$. That is, the states
created by the operator $\Phi _{-1-j,\mp m,\mp \overline{m}}^{\pm \omega }$
have the same conformal dimension than those created by $\Phi _{j,m,%
\overline{m}}^{\omega }$.

Operators (\ref{vertex}) have the following OPE with the $sl(2)_{k}$
Kac-Moody currents%
\begin{eqnarray}
J^{3}(z)\Phi _{j,m,\overline{m}}^{\omega }(w) &\simeq &\frac{(m+k\omega /2)}{%
(z-w)}\Phi _{j,m,\overline{m}}^{\omega }(w)+\mathcal{O}(1)  \label{j3} \\
J^{\pm }(z)\Phi _{j,m,\overline{m}}^{\omega }(w) &\simeq &\frac{(\pm j-m)}{%
(z-w)^{1\pm \omega }}\Phi _{j,m\pm 1,\overline{m}}^{\omega }(w)+\mathcal{O}%
((z-w)^{\mp \omega })  \label{jpm}
\end{eqnarray}

The theory also admits conjugate representations of the vertex operators.
These are important ingredients in our discussion. Such conjugate
representations are defined by operators

\begin{equation}
\widetilde{\Phi }_{j,m,\overline{m}}^{\omega }(z)=\frac{1}{Z_{j,m}}\beta
_{(z)}^{j+m}e^{-\sqrt{\frac{2}{k-2}}(j+\frac{k}{2})\phi (z)}e^{i\sqrt{\frac{2%
}{k}}(m-\frac{k}{2})X(z)}e^{i\sqrt{\frac{2}{k}}(m+\frac{k}{2}\omega
)T(z)}\times h.c.  \label{conjugated}
\end{equation}%
which create conjugate \textit{in}-states%
\begin{equation}
\lim_{z\rightarrow 0}\widetilde{\Phi }_{j,m,\overline{m}}^{\omega
}(z)\left\vert 0\right\rangle =\left\vert j_{n},m_{n},\overline{m}%
_{n},\omega _{n}\right) .  \label{inconjugated}
\end{equation}

In (\ref{conjugated}) $Z_{j,m}^{-1}$ stands for a normalization factor,
which gets fixed once one requires the two-point function between an
operator (\ref{vertex}) and its conjugate (\ref{conjugated}) to be
normalized to one; namely $\left\langle j,m,\overline{m},\omega \right.
\left\vert j,-m,-\overline{m},-\omega \right) \equiv 1$. This yields

\begin{equation}
Z_{j,m}=(-1)^{j+m}c_{0}\ \Gamma (j+m+1).  \label{Z}
\end{equation}

Conjugate representation (\ref{conjugated}) was introduced in \cite%
{GiribetLopez}. These operators can be thought of as a twisted version of
the operators proposed in \cite{BK} to describe discrete states in the
two-dimensional black hole background. Operators (\ref{conjugated}) create
states in a conjugate representations $\left\vert j_{n},m_{n},\overline{m}%
_{n},\omega _{n}\right) $. This is analogous to the $SU(2)$ case studied in 
\cite{Dotsenko1}.

It is easy to verify that operators $\Phi _{j,m,\overline{m}}^{\omega }$ and 
$\widetilde{\Phi }_{j,m,\overline{m}}^{\omega }$ create states with the same
conformal dimension (\ref{h}). Besides, one can also verify that (\ref%
{conjugated}) obeys the following OPE with the currents

\begin{eqnarray}
J^{3}(z)\widetilde{\Phi }_{j,m,\overline{m}}^{\omega }(w) &\simeq &\frac{%
(m+k\omega /2)}{(z-w)}\widetilde{\Phi }_{j,m,\overline{m}}^{\omega }(w)+%
\mathcal{O}(1) \\
J^{\pm }(z)\widetilde{\Phi }_{j,m,\overline{m}}^{\omega }(w) &\simeq &\frac{%
(\mp 1\mp j-m)}{(z-w)^{1\pm \omega }}\widetilde{\Phi }_{j,m\pm 1,\overline{m}%
}^{\omega }(w)+\mathcal{O}((z-w)^{\mp \omega });
\end{eqnarray}%
that is, conjugate operators $\widetilde{\Phi }_{j,m,\overline{m}}^{\omega }$
have exactly the same properties that the Weyl-reflected operator $\Phi
_{-1-j,m,\overline{m}}^{\omega }$. In fact, as pointed out in \cite%
{GiribetLopez}, Weyl reflection can also be thought of as a conjugation
operation associated to the zero-dimension field $\Phi
_{-1,0,0}^{0}(z)=\gamma _{(z)}^{-1}e^{-\sqrt{\frac{2}{k-2}}\phi (z)}\times
h.c.$

Important ingredients of the Coulomb gas realization that the free field
approach leads to are the screening operators. These are given by%
\begin{equation}
\widetilde{\Phi }_{1-\frac{k}{2},\frac{k}{2},\frac{k}{2}}^{-1}(z)=\beta (z)\
e^{-\sqrt{\frac{2}{k-2}}\phi (z)}\times h.c.  \label{screening}
\end{equation}%
These operators have conformal dimension one and regular OPE with the
Kac-Moody currents.

Another special case of operators (\ref{conjugated}) is the conjugate
representation of the identity operator. This is given by the zero-dimension
field%
\begin{equation}
\widetilde{\Phi }_{0,0,0}^{0}(z)=\Phi _{-\frac{k}{2},-\frac{k}{2},-\frac{k}{2%
}}^{1}(z)=e^{-\sqrt{\frac{2}{k-2}}\frac{k}{2}\phi (z)}e^{-i\sqrt{\frac{k}{2}}%
X(z)}\times h.c.  \label{identity}
\end{equation}

Operator (\ref{identity}) was first introduced by Fateev and the brothers
Zamolodchikov in their renowned FZZ unpublished paper \cite{FZZ}, and in
reference \cite{MO3} it was dubbed \textit{spectral flow operator}.
Representation (\ref{identity})\ is important to define the charge asymmetry
conditions; see \cite{Dotsenko1} for the details.

\section{String scattering amplitudes in AdS$_{3}$}

In references \cite{GN2,GN3,GiribetLopez}, conjugate representations were
considered to describe string scattering amplitudes in AdS$_{3}$ in the case
where the winding number is taken into account. Based on an adaptation of
Dotsenko works \cite{Dotsenko1, Dotsenko2}, a prescription was proposed to
calculate the correlators on $\mathbb{H}_{+}^{3}/U(1)\times \mathbb{R}$.
According to such prescription, the string scattering amplitudes of $n$%
-strings in AdS$_{3}$ are obtained by integrating over the worldsheet the
following correlation function%
\begin{equation}
X_{n}^{\Delta \omega }=\ \left\langle j_{1},m_{1},\overline{m}_{1},\omega
_{1}\right\vert \tprod\nolimits_{t=2}^{p}\Phi _{j_{t},m_{t},\overline{m}%
_{t}}^{\omega _{t}}(z_{t})\tprod\nolimits_{l=p+1}^{n-1}\widetilde{\Phi }%
_{j_{l},m_{l},\overline{m}_{l}}^{\omega _{l}}(z_{l})\left\vert j_{n},m_{n},%
\overline{m}_{n},\omega _{n}\right)  \label{correlationfunction}
\end{equation}%
where $\Delta \omega =\tsum\nolimits_{i=1}^{n}\omega _{i}=p+1-n$ (notice
that $p\geq 1$.) That is, the tree-level string amplitude is given by 
\begin{equation}
\mathcal{A}_{\text{string}}^{\Delta \omega }=\int
\tprod\nolimits_{l=3}^{n-1}d^{2}z_{l}\ X_{n}^{\Delta \omega },
\label{stringtheory}
\end{equation}%
integrating over $n-3$ vertex insertions on the sphere.

Here we are concerned with the amplitudes of processes in which the total
winding number is violated in $n-2$ units; namely, we will consider
correlation functions%
\begin{equation}
X_{n}^{2-n}=\left\langle j_{1},m_{1},\overline{m}_{1},\omega _{1}\right\vert
\tprod\nolimits_{l=2}^{n-1}\widetilde{\Phi }_{j_{l},m_{l},\overline{m}%
_{l}}^{\omega _{l}}(z_{l})\left\vert j_{n},m_{n},\overline{m}_{n},\omega
_{n}\right) .  \label{loquequiero}
\end{equation}

For this correlator not to vanish, it is necessary to insert a precise
amount $s$ of screening operators (\ref{screening}). $s$ is determined by
the charge asymmetry condition corresponding to the field $\phi (z)$, which
yields $s=-1-\tsum\nolimits_{i=1}^{n}j_{i}-(n-2)k/2$. On the other hand, the
charge asymmetry conditions corresponding to the fields $X(z)$ and $T(z)$
demand $\tsum\nolimits_{i=1}^{n}m_{i}=\tsum\nolimits_{i=1}^{n}\overline{m}%
_{i}=(n-2)k/2$ and $\tsum\nolimits_{i=1}^{n}\omega _{i}=2-n$.

For further purpose it will be necessary to renormalize the vertex operators 
$\widetilde{\Phi }_{j_{i},m_{i},\overline{m}_{i}}^{\omega _{i}}$ of the $n-2$
intermediate states, $i=2,3,4,...n-1$. To do so, first we rewrite $Z_{j,%
\overline{m}}$ as follows%
\begin{equation*}
Z_{j,\overline{m}}=(-1)^{j+m}c_{0}\ \Gamma (j+\overline{m}%
+1)=\lim_{\varepsilon \rightarrow 0}\text{ }c_{0}\ Z_{j,\overline{m}%
}^{(\varepsilon )}\qquad \text{with}\qquad Z_{j,\overline{m}}^{(\varepsilon
)}=\frac{\Gamma (\varepsilon )}{\Gamma (\varepsilon -j-\overline{m})},
\end{equation*}%
and then introduce a regularization factor to extract the divergence by
renormalizing $c_{0}$; namely%
\begin{equation*}
\tprod\nolimits_{l=2}^{n-1}\frac{1}{Z_{j,m}Z_{j,\overline{m}}}%
=\lim_{\varepsilon \rightarrow 0}\varepsilon ^{2-n}\ (c_{0}/\varepsilon
)^{2-n}\tprod\nolimits_{l=2}^{n-1}\frac{1}{Z_{j,m}Z_{j,\overline{m}%
}^{(\varepsilon )}}=c^{2-n}\tprod\nolimits_{l=2}^{n-1}(-1)^{-j_{l}-m_{l}}%
\frac{\Gamma (-j_{l}-\overline{m}_{l})}{\Gamma (1+j_{l}+m_{l})}.
\end{equation*}

The amplitudes of a scattering process of $n$ strings in which the winding
number conservation is violated in $n-2$ units is then given by integrating
the correlation function%
\begin{eqnarray*}
X_{n}^{2-n} &=&\frac{(-1)^{s-2j_{n}-m_{n}-\overline{m}_{n}}\Gamma (-s)c^{2-n}%
}{\Gamma (1+j_{n}+m_{n})\Gamma (1+j_{n}+\overline{m}_{n})}%
\tprod\nolimits_{l=2}^{n-1}\frac{(-1)^{-j_{l}-m_{l}}\Gamma (-j_{l}-\overline{%
m}_{l})}{\Gamma (1+j_{l}+m_{l})} \\
&&\int \tprod\nolimits_{r=1}^{s}d^{2}y_{r}\left\langle \gamma
_{(z_{1})}^{-1-j_{1}-m_{1}}\tprod\nolimits_{l=2}^{n}\beta
_{(z_{l})}^{j_{l}+m_{l}}\tprod\nolimits_{r=1}^{s}\beta
_{(y_{r})}\right\rangle
\end{eqnarray*}%
\begin{equation*}
\left\langle e^{-\sqrt{\frac{2}{k-2}}(j_{1}+1)\phi
(z_{1})}\tprod\nolimits_{l=2}^{n}e^{-\sqrt{\frac{2}{k-2}}(j_{l}+\frac{k}{2}%
)\phi (z_{l})}\tprod\nolimits_{r=1}^{s}e^{-\sqrt{\frac{2}{k-2}}\phi
(y_{r})}\right\rangle
\end{equation*}%
\begin{equation*}
\left\langle e^{i\sqrt{\frac{2}{k}}m_{1}X(z_{1})}\tprod%
\nolimits_{l=2}^{n}e^{i\sqrt{\frac{2}{k}}(m_{l}-\frac{k}{2}%
)X(z_{l})}\right\rangle \left\langle e^{i\sqrt{\frac{2}{k}}(m_{1}+\frac{k}{2}%
\omega _{1})T(z_{1})}\tprod\nolimits_{l=2}^{n}e^{i\sqrt{\frac{2}{k}}(m_{l}+%
\frac{k}{2}\omega _{l})T(z_{l})}\right\rangle \times h.c.
\end{equation*}%
where the integrals over $y_{r}$ come from the insertion of $s$ screening
operators, with $s=-1-\tsum\nolimits_{i=1}^{n}j_{i}-(n-2)k/2$; and where we
set $z_{1}=\infty $, $z_{2}=1$, and $z_{n}=0$.

Expanding the Wick contractions, and considering the free field propagators (%
\ref{propagators}), one finds the integral expression%
\begin{equation*}
X_{n}^{2-n}=c^{2-n}\tprod\nolimits_{i=1}^{n}\frac{\Gamma (-j_{i}-\overline{m}%
_{i})}{\Gamma (1+j_{i}+m_{i})}\tprod\nolimits_{i<j}^{n-1,n}(z_{i}-z_{j})^{%
\beta _{ij}}(\overline{z}_{i}-\overline{z}_{j})^{\overline{\beta }_{ij}}
\end{equation*}%
\begin{equation}
\tprod\nolimits_{i<j}^{n-1,n}|z_{i}-z_{j}|^{-2\alpha _{i}\alpha _{j}}\Gamma
(-s)\int
\tprod\nolimits_{r=1}^{s}d^{2}y_{r}\tprod\nolimits_{i=1}^{n}\tprod%
\nolimits_{r=1}^{s}|z_{i}-y_{r}|^{-2\alpha
_{i}b}\tprod\nolimits_{r<t}^{s-1,s}|y_{r}-y_{t}|^{-2b^{2}},  \label{Uh}
\end{equation}%
where we introduced the notation $\alpha _{i}=b(j_{i}+1+b^{-2}/2)$ with $%
b^{-2}=k-2$, and $\beta _{ij}=k/2-m_{i}-m_{j}-k\omega _{i}\omega
_{j}/2-m_{i}\omega _{j}-m_{j}\omega _{i}$, and analogously for $\overline{%
\beta }_{ij}$ changing $m_{i}$ and $\omega _{i}$ for $\overline{m}_{i}$ and $%
\overline{\omega }_{i}$ respectively. Notice that, in terms of $\alpha _{i}$%
, we have $s=b^{-1}\tsum\nolimits_{i=1}^{n}\alpha _{i}+1+b^{-2}$.

In finding an expression like (\ref{Uh}), the rapid way of dealing with the
contraction of the $\beta $-$\gamma $ system is that of first assuming the
case of $j_{1}-m_{1}$ being a positive integer and then extending the
resulting expressions. Also, it was used in (\ref{Uh}) that the product of
the multiplicity factor coming from the Wick contraction of the $\beta $-$%
\gamma $ contribution and the normalization of the $n^{\text{th}}$ vertex
can be rewritten as%
\begin{equation*}
\frac{\Gamma (-j_{1}-m_{1})\Gamma (-j_{1}-\overline{m}_{1})}{\Gamma
(1+j_{n}+m_{n})\Gamma (1+j_{n}+\overline{m}_{n})}=\frac{\Gamma (-j_{1}-%
\overline{m}_{1})}{\Gamma (1+j_{1}+m_{1})}\frac{\Gamma (-j_{n}-\overline{m}%
_{n})}{\Gamma (1+j_{n}+m_{n})}(-1)^{j_{n}-j_{1}+\overline{m}_{n}-m_{1}}.
\end{equation*}

The $z_{i}$-dependent factor in the first line of (\ref{Uh}) comes from the
Wick contraction of the fields $X(z)$ and $T(z)$. In the second line of (\ref%
{Uh}), on the other hand, one already sees the $n$-point Liouville
correlation function appearing. In fact, Liouville correlation functions of
primary operators $V_{\alpha _{i}}(z_{i})=e^{\sqrt{2}\alpha _{i}\phi
(z_{i})} $ are given by%
\begin{equation*}
\left\langle \tprod\nolimits_{i=1}^{n}V_{\alpha _{i}}(z_{i})\right\rangle _{%
\text{L}}=\int \mathcal{D\varphi }\ e^{-\frac{1}{4\pi }\int d^{2}w((\partial 
\mathcal{\varphi })^{2}+(b+1/b)R\mathcal{\varphi }+4\pi e^{2b\mathcal{%
\varphi }})}\tprod\nolimits_{i=1}^{n}e^{\sqrt{2}\alpha _{p}\mathcal{\varphi }%
(z_{i})}=
\end{equation*}%
\begin{equation*}
=\frac{\Gamma (-s)}{b}\tprod\nolimits_{i<j}^{n-1,n}|z_{i}-z_{j}|^{-2\alpha
_{i}\alpha _{j}}\int
\tprod\nolimits_{p=1}^{s}d^{2}y_{p}\tprod\nolimits_{i=1}^{n}\tprod%
\nolimits_{l=1}^{s}|z_{i}-y_{l}|^{-2\alpha
_{i}b}\tprod\nolimits_{l<t}^{s-1,s}|y_{l}-y_{t}|^{-2b^{2}}
\end{equation*}%
with $s=-b^{-1}\tsum\nolimits_{i=1}^{n}\alpha _{i}+1+b^{-2}$. This means
that, after absorbing an irrelevant factor, we can write correlation
functions (\ref{loquequiero}) as follows%
\begin{equation}
X_{n}^{2-n}=c^{2-n}\tprod\nolimits_{i=1}^{n}\frac{\Gamma (-j_{i}-\overline{m}%
_{i})}{\Gamma (1+j_{i}+m_{i})}\tprod\nolimits_{l<t}^{n-1,n}(z_{l}-z_{t})^{%
\beta _{lt}}(\overline{z}_{l}-\overline{z}_{t})^{\overline{\beta }_{lt}}\
\left\langle \tprod\nolimits_{i=1}^{n}V_{\alpha _{i}}(z_{i})\right\rangle _{%
\text{L}},  \label{final}
\end{equation}%
recalling $\beta _{lt}=k/2-m_{l}-m_{t}-k\omega _{l}\omega _{t}/2-m_{l}\omega
_{t}-m_{t}\omega _{l}$, $\alpha _{i}=b(j_{i}+1+b^{-2}/2)$, and $b^{-2}=k-2$.

\section{Conclusions}

Expression (\ref{final}) is exactly the expression conjectured in \cite{R}
for the case $|\Delta \omega |=n-2$, and this is what we wanted to prove.
The worldsheet conformal field theory calculation in terms of free fields
actually confirms that formula. It is worth pointing out that resorting to
the prescription of \cite{GN2, GN3, GiribetLopez} in terms of conjugate
representations was crucial to find (\ref{final}); a free field calculation
in terms of the standard Wakimoto representation for the vertices (c.f. \cite%
{Becker}) would never lead to such a direct computation, in particular
because it is not clear in that case how to implement the winding number
violation. Therefore, the result obtained here can be interpreted as a
non-trivial test passed by the prescription of \cite{GN2, GN3, GiribetLopez}%
, which seems to be powerful enough to yield an expression like (\ref{final}%
) even in a case in which the modular differential equations are not at
hand. Of course, even when convincing, a computation based on a free field
realization can hardly be considered a rigorous proof; in particular, it
strongly relies on analytic continuation of the integral formulas involved.
However, it is still interesting that formula (\ref{final}) is confirmed by
these means. It has already been argued in \cite{R} that free field
computations in the FZZ dual theory done by Fateev in an unpublished paper 
\cite{Fateev} gave further evidence in favor of the validity of (\ref{final}%
).%
\begin{equation*}
\end{equation*}%
This work was supported by ANPCyT, CONICET,\ and UBA.

\section{Addendum}

Let us revisit the computation of maximally winding violating string amplitudes in three-dimensional Anti-de Sitter space discussed above. Here, we give an alternative derivation of these observables. This derivation, the simplest to the best of our knowledge, follows from identities between spectrally flowed representations of $\hat{sl}(2)_k$ Kac-Moody algebra and, in contrast to the one described above, it does not resort to conjugate representations with auxiliary fields, but it rather involves the standard Wakimoto free field representation.   

The generators of the Kac-Moody $\hat{sl}(2)_k$ algera satisfy the Lie products
\begin{equation}
[J_n^3 , J^{\pm }_m] = \pm  J^{\pm }_{n+m} \ , \ \ \ [J_n^3 , J^3_m] =  \frac{k}{2} m\delta_{n+m,0} \ , \ \ \ [J_n^+ , J^{- }_m] = -2  J^{3 }_{n+m}+kn\delta_{n+m,0} \ . \label{ERP}
\end{equation}
with $a=3,\pm $. These brackets, and its complex conjugate counterpart, are realized by defining the local currents
\begin{equation}
J^a(z)=\sum_{n\in \mathbb{Z}} J_n \ z^{-1-n} \ , \ \ \ \bar{J}^a(\bar{z})=\sum_{n\in \mathbb{Z}} \bar{J}_n \ \bar{z}^{-1-n}
\end{equation}
and computing the operator product expansion (OPE) among them. A useful representation of these local currents has been given by Wakimoto \cite{Wakimoto}, who proposed
\begin{eqnarray}
J^{+}(z) &=&\beta (z), \\
J^{3}(z) &=&-\beta (z)\gamma (z)-\sqrt{\frac{k-2}{2}}\partial \phi (z), \\
J^{-}(z) &=&\beta (z)\gamma ^{2}(z)+\sqrt{2k-4}\gamma (z)\partial \phi
(z)+k\partial \gamma (z).
\end{eqnarray}%
with the free field propagators%
\begin{equation}
\left\langle \phi (z)\phi (w)\right\rangle =-\log (z-w)\text{%
,\quad } \ \ \left\langle \beta (z)\gamma (w)\right\rangle =\frac{1}{(z-w)}\text{;%
}  \label{propagators}
\end{equation}%
and analogously for the anti-holomorphic contributions.

Algebra (\ref{ERP}) is invariant under the spectral flow operation
\begin{equation}
J^3_n \to  {J}^3_{n}+ \frac{k}{2}\omega \delta_{n,0} \ , \ \ \ \ \ \ \ J^{\pm}_n \to  {J}^{\pm}_{n\pm \omega } . \label{SP}
\end{equation}
This generates a whole family of new representations, usually denoted $\mathcal{C}^{\alpha ,\omega }_j$ and $\mathcal{D}^{\pm ,\omega}_j$, and hereafter called spectrally flowed representations. More precisely, for $|\omega |>1$, automorphism (\ref{SP}) does generate new representations; however, the cases $\omega = \pm 1$ are special in the sense that the highest (and lowest) weight representations of the sector $\omega = 0 $ coincide with lowest (resp highest) weight representations of the sector $\omega = -1$ (resp. $\omega = +1$). This results in the identification of the discrete representations
\begin{equation}
\mathcal{D}_j^{\pm , \omega = 0} \ \leftrightarrow \ \mathcal{D}_{-\frac{k}{2}-j}^{\mp , \omega = \pm 1} \ ,\label{dualityAA}
\end{equation} 
which will be crucial for the argument herein.

The new Kac-Moody primaries $\left\vert j,m,\omega \right\rangle$, which are annihilated by the positive modes of the new (spectrally flowed) currents, are essential to construct the string spectrum \cite{MO1}. Algebraically, these states are defined as those that obey 
\begin{equation}
J^3_0 \left\vert j,m,\omega \right\rangle = \left(m+\frac{k}{2}\omega \right)\left\vert j,m,\omega \right\rangle \ , \ \ \ \bar{J}^3_0 \left\vert j,\bar{m},\omega \right\rangle = \left(\bar{m}+\frac{k}{2}\omega \right)\left\vert j,\bar{m},\omega \right\rangle \ , \label{9AA}
\end{equation}
together with 
\begin{equation}
J^{\pm}_{n>\mp \omega } \left\vert j,m,\omega \right\rangle = 0 \ , \ \ \ \ \ \bar{J}^{\pm}_{n>\mp \omega } \left\vert j,\bar{m},\omega \right\rangle = 0 \ . \label{9AAA}
\end{equation}

In the case of long strings, corresponding to states of the continuous representations $\mathcal{C}^{\alpha ,\omega }_j$, the parameter $\omega $ of the spectral flow transformation is interpreted as the winding number of the asymptotic states, associated to the presence of a non-vanishing $B$-field in the background. In the case short strings, those described by states of the discrete representations $\mathcal{D}^{\pm ,\omega }_j$, the geometrical interpretation of $\omega $ is less clear, but it still contributes to the mass-shell condition in a way that resembles a winding number.  

Due to the duality among different representations (\ref{dualityAA}), it will be enough for our purpose to consider the spectral flow sector $\omega =0$. In terms of the Wakimoto fields, the vertex operators that create the states of this sector take the form
\begin{equation}
\Phi _{j,m,\bar{m}}^{\omega =0}(z)=c_{0}\ \gamma^{j-m} (z) \bar{\gamma}^{j-\bar{m}}(\bar{z}) \ e^{\sqrt{%
\frac{2}{k-2}}j\phi (z,\bar{z})} , \label{WakiAA}
\end{equation}%
where $c_0$ is a normalization constant that here we will set to 1 for convention. It can be easily checked that operators (\ref{WakiAA}) of the spectral flow sector $\omega =0 $ have the following OPE with the $sl(2)_{k}$
Kac-Moody currents%
\begin{eqnarray}
J^{3}(z)\Phi _{j,m,\bar{m}}^{0}(w) &\simeq &\frac{m}{%
(z-w)}\Phi _{j,m,\bar{m}}^{0 }(w)+\ ...  \label{j3} \\
J^{\pm }(z)\Phi _{j,m,\bar{m}}^{0 }(w) &\simeq &\frac{(\pm j-m)}{%
(z-w)}\Phi _{j,m\pm 1,\bar{m}}^{0 }(w)+\ ...   \label{jpmAA}
\end{eqnarray}
which realize (\ref{9AA})-(\ref{9AAA}) for $\omega =0$.

Vertex operators $\Phi_{j,m,\bar{m}}^{\omega }$ are the objects that create the states $\vert j,m,\omega \rangle \times \vert j,\bar{m},\omega \rangle$ out of the $SL(2,\mathbb{R})\times SL(2,\mathbb{R})$ invariant vacuum $\left\vert 0\right\rangle $; namely%
\begin{equation}
\lim_{z\rightarrow 0}\Phi _{j,m,\bar{m}}^{\omega }(z,\bar{z})\left\vert
0\right\rangle =\left\vert j,m,\omega \right\rangle \times \left\vert j,\bar{m},\omega \right\rangle  \label{inA}
\end{equation}%

The conformal dimension of these states is given by%
\begin{equation}
h_{j,m}^{\omega }=-\frac{j(j+1)}{k-2}-m\omega -\frac{k}{4}\omega ^{2} \ , \ \ \ h_{j,\bar{m}}^{\omega }=-\frac{j(j+1)}{k-2}-\bar{m}\omega -\frac{k}{4}\omega ^{2}
\label{hAA}
\end{equation}

Notice that, indeed, the states $\left\vert j,\pm j ,0 \right\rangle$ and $\left\vert -{k}/{2}-j,\pm {k}/{2}\pm j ,\mp 1 \right\rangle$ have the same quantum numbers, in accordance to (\ref{dualityAA}). That is, they have the same eigenvalue under $J^3_0$, namely $\pm j$, and the same conformal weight $h_{j,\pm j}^{0 }=h_{-{k}/{2}-j,\pm {k}/{2}\pm j}^{\mp 1 }$. The formula for the conformal weight (\ref{hAA}) also remains
unchanged under the Weyl reflection $j\rightarrow -1-j$. That is, the states $\left\vert j,m ,\omega \right\rangle$ and $\left\vert -1-j,m ,\omega \right\rangle$ have the same quantum numbers, and in particular $h^{\omega}_{j,m}=h^{\pm\omega}_{-1-j,\pm m}$. 

Consider now the maximally winding violating correlation function
\begin{equation}
X_n^{2-n}=\Big\langle \prod_{a=1}^{2}\Phi_{-1-j_a , m_a , m_a}^{0 }(z_a,\bar{z}_a) \prod_{i=3}^{n} \Phi_{j_i , j_i , j_i}^{+1 }(z_i,\bar{z}_i)\Big\rangle
\end{equation}
which involves $n-2$ highest-weight states of the spectral flow sector $\omega = 1$ (the argument works for $\omega = -1$ as well). In virtue of (\ref{dualityAA}) we can equal this correlator to the following one
\begin{equation}
X_n^{2-n}={c}^{2-n}\Big\langle \prod_{a=1}^{2}\Phi_{-1-j_a , m_a , m_a}^{0 }(z_a,\bar{z}_a)  \prod_{i=3}^{n} \Phi_{-\frac{k}{2}-j_i , \frac{k}{2}+j_i , \frac{k}{2}+j_i}^{0 }(z_i,\bar{z}_i)\Big\rangle\label{16AA}
\end{equation}
which only involves vertices (\ref{WakiAA}), of the sector $\omega =0$. This is different to the computation done above, where the presence of vertices with $\omega \neq 0$ demands the inclusion of extra fields. The factor $c^{2-n}$ in (\ref{16AA}) stands for the relative normalization between the operators $\Phi_{j,\pm j,\pm j}^{0}$ and $\Phi_{-\frac{k}{2}-j,\pm\frac{k}{2}\pm j,\pm\frac{k}{2}\pm j}^{\mp 1}$, which remains unspecified \cite{R}.

Involving only operators of the unflowed sector, correlator (\ref{16AA}) admits to be computed by using the Wakimoto representation (\ref{WakiAA}), straightforwardly applying the techniques described in reference \cite{Becker}, with no need of auxiliary fields cf. \cite{GN2,GN3}. This results in
\begin{eqnarray}
X_n^{2-n}&=&c^{2-n}\prod_{a=1}^{2}\frac{\Gamma(-j_a-m_a)}{\Gamma(1+j_a+m_a)} \ \prod_{i<j}^n|z_i-z_j|^{-\frac{4}{k-2}(j_i+\frac{k}{2})(j_j+\frac{k}{2})+2b_{ij}} \times  \nonumber \\
&&\Gamma(-s) \int \prod_{r=1}^{s}d^2w_r \ \prod_{r=1}^{s}\prod_{i=1}^{n} |z_i-w_r|^{-\frac{4}{k-2}(j_i+\frac{k}{2})}
\prod_{r<t}^{s} |w_t-w_r|^{-\frac{4}{k-2}}
 , \label{ESTAA}
\end{eqnarray}
where
\begin{equation}
s=-1-\sum_{i=1}^n j_i -\frac{k}{2} (n-2)
\end{equation}
and $b_{ij}=0$ for $i,j>3,4,... n$; $b_{ai}=j_a$ for $a=1,2$ and $i>2$; $b_{ab}=j_a + j_b -2b^2$ for $a,b=1,2$.

The integrand in (\ref{ESTAA}) follows from the operator product expansions
\begin{eqnarray}
\gamma(z_a)^{j_a-m_a}\bar{\gamma}(\bar{z}_a)^{j_a-m_a}\ \prod_{r=1}^s \beta(w_r)\bar{\beta}(\bar{w}_r)  \ \simeq  \frac{\Gamma^2(1+j_a+m_a+s)}{\Gamma^2(1+j_a+m_a)} \prod_{r=1}^s|z_a-w_r|^{-2}  + \ ...  \nonumber 
\end{eqnarray}
and
\begin{eqnarray}
&&e^{-\sqrt{\frac{2}{k-2}}(j_a+1)\phi (z_a,\bar{z}_a)}  \ \prod_{r=1}^s e^{-\sqrt{\frac{2}{k-2}}\phi(w_r,\bar{w}_r)}\ \simeq  \prod_{r=1}^s|z_a-w_r|^{-\frac{4}{k-2}(j_a+1)} + \ ...  \ , \nonumber \\
&&e^{-\sqrt{\frac{2}{k-2}}(j_i+\frac{k}{2})\phi (z_i,\bar{z}_i)}  \ \prod_{r=1}^s e^{-\sqrt{\frac{2}{k-2}}\phi(w_r,\bar{w}_r)}\ \simeq  \prod_{r=1}^s|z_i-w_r|^{-\frac{4}{k-2}(j_i+\frac{k}{2})} + \ ...  \ , \nonumber 
\end{eqnarray}
where the ellipses stand for subleading contributions with less Wick contractions. Recall that the $\beta $-dependent operators 
\begin{equation}
\int d^2w\ \beta(w)\bar{\beta}(\bar{w}) e^{-\sqrt{\frac{2}{k-2}}\phi(w,\bar{w})}
\end{equation}
come from the interaction term of the $SL(2,\mathbb{R})$ WZW action when written in the Wakimoto representation \cite{Becker} and, in the Coulomb gas approach, they act as certain amount ($s$) of screening operators needed to compensate the dilatonic background charge.

The integral on the right hand side of (\ref{ESTAA}) can also be identified as a correlation function in Liouville field theory. More precisely, the $n$-point function of exponential primary operators in Liouville theory takes the form \cite{GL}
\begin{equation}
\Big\langle \prod_{i=1}^n V_{\alpha_i} (z_i,\bar{z}_i)\Big\rangle _{\text{L}} =  \Gamma(-s) \prod_{i<j}^n|z_i-z_j|^{-4\alpha_i\alpha_j} \int \prod_{r=1}^{s}d^2w_r \ \prod_{r=1}^{s}\prod_{i=1}^{n} |z_i-w_r|^{-{4}b\alpha_i} \prod_{r<t}^{s} |w_t-w_r|^{-4b^2}\label{OchoA}
\end{equation}
with
\begin{equation}
bs+\sum_{i=1}^n\alpha_i =Q \ , \ \ \ \ \ Q=b+\frac{1}{b} \ .
\end{equation}
In these variables, the Liouville central charge reads $c=1+6Q^2$. Then, the dictionary between (\ref{ESTAA}) and (\ref{OchoA}) is simple and is the one of \cite{R}; namely
\begin{equation}
\alpha_i = b\left( j_i + \frac{b^2}{2} + 1\right) \ , \ \ \ \ \ b^2=\frac{1}{k-2} \ .
\end{equation}

In conclusion, we arrive to the formula
\begin{equation}
X_n^{2-n}=c^{2-n} \ \prod_{i=1}^{2}\frac{\Gamma(-j_i-m_i)}{\Gamma(1+j_i+m_i)} \prod_{i<j}^{n}|z_i - z_j|^{2b_{ij}} \Big\langle \prod_{i=1}^n V_{\alpha_i} (z_i,\bar{z}_i)\Big\rangle _{\text{L}} ,\label{FinalG}
\end{equation}
which is analogous to the one conjectured in \cite{R} and what we actually wanted to prove. 

In conclusion: We have derived formula (\ref{FinalG}), which expresses the $n$-point function of maximally winding violating processes in AdS$_3$ in terms of $n$-point correlation functions of Liouville field theory. This is analogous to the expressions proposed in \cite{R}, here obtained in a remarkably succinct way without resorting to nothing but well-known dualities among spectrally flowed representations and to the standard Wakimoto fields, with no need of auxiliary fields cf. \cite{GN2,GN3}. 

Despite its simplicity, the derivation presented here has to be regarded as complementary to that presented in the previous sections, and by no means as its generalization. This is because the one here has its limitations as well: it only involves states with winding numbers $\omega =0,\pm 1$ and deals with the cases in which the $n-2$ states with $\omega \neq 0$ belong to the highest or lowest weight representations. Still, this is the simplest derivation of winding violating processes in AdS$_3$ and the simplest example of WZW-Liouville correspondence given so far.

\end{document}